\documentclass[amssymb,aps,pra,byrevtex,superscriptaddress,twocolumn]{revtex4}
\usepackage{epsfig,graphics,graphicx}
\usepackage{amssymb,amsmath,amsthm}
\usepackage{verbatim}
\usepackage{ulem}
\usepackage{cancel}
\usepackage[colorlinks=true,linkcolor=blue,citecolor=blue]{hyperref}

\begin{document}
\title{Localized chaos of elliptically polarized cavity solitons in broad-area VCSEL with saturable absorber}

\author{K. Panajotov}
\affiliation{Department of Applied Physics and Photonics, B-Phot, Vrije Universiteit Brussel (VUB), Pleinlaan 2, B-1050 Brussels, Belgium }
\affiliation{Institute of Solid State Physics, 72 Tzarigradsko Chaussee Blvd., 1784 Sofia, Bulgaria}
\email{kpanajot@b-phot.org}
\author{M. Tlidi}
\affiliation{Facult{\'e} des Sciences, Universit{\'e} Libre de Bruxelles (U.L.B.), C.P.2 31, Campus Plaine, B-1050 Bruxelles, Belgium}

\begin{abstract}
We introduce a spin-flip model for a broad-area Vertical-Cavity Surface-Emitting Laser (VCSEL) with a saturable absorber. We demonstrate simultaneous existence of orthogonally linearly polarized and elliptically polarized cavity solitons. We show that polarization degree of freedom leads to period doubling route to spatially localized chaos of the elliptically polarized cavity solitons.
\end{abstract}

\maketitle

Polarization dynamics has attracted considerable interest for both solitary Vertical-Cavity Surface-Emitting Lasers (VCSELs) \cite{Choquette_apl94,Regalado_jqe97,Panajotov_aip01,Nagler_pra03,Verschaffelt_oer01} and for lasers subject to optical injection \cite{Pan_apl93, Ryvkin_jap04, Gatare_pra07, Panajotov_jqe09} and optical feedback \cite{Arteaga_jqe06, Panajotov_apl04}; for reviews see \cite{Panajotov_ch13, Panajotov_jstqe13}. Such dynamics appears because of the lack of strong polarization selectivity mechanism in VCSELs: the cylindrical symmetry removes the waveguiding and the reflectivity anisotropy and the surface emission removes the gain anisotropy \cite{Panajotov_ch13}. Although many experiments can be explained by simple rate equation model with self and cross gain saturation (see e.g. \cite{Nagler_pra03,Ryvkin_jap04}), the emission of elliptically polarized light can only be explained by the so called Spin-Flip Model  or SFM \cite{SanMiguel_pra95}. SFM considers transitions between conduction band and heavy hole valence band in quantum well active material as two sub-systems of different spin orientation that are coupled through the spin-flip processes. Initially, SFM has explained polarization switching \cite{Regalado_jqe97,Susanto_pra15} and more recently, the appearance of deterministic polarization  chaos in solitary VCSELs \cite{Virte_np13,Raddo_sr17}.

Accounting for the light polarization also becomes important when considering Cavity Solitons (CSs) in broad area VCSELs. CS are localized light structures in the transverse plane of externally driven nonlinear resonators (for reviews see \cite{Lugiato_po94,Rosanov_po96,Staulinas_book,Ackemann_aamoo09,Tlidi_ptrsa14,Tlidi_ptrsa18}). VCSELs are very attractive for CS studies because of the large Fresnel number possible, the short cavity and their mature technology \cite{Taranenko_pra00,Hachair_jstqe06}. First experimental indication for the importance of light polarization has been obtained for 40 $\mu m$-diameter VCSELs \cite{Hachair_pra09}. Recently, vector CSs have been observed in 80 $\mu m$ - diameter VCSEL \cite{Averlant_sr16} and explained by the SFM model. Two orthogonal polarized vector solitons, both Gaussian and ring shaped, have been experimentally demonstrated for broad-area VCSEL with frequency-selective feedback in \cite{Rodriguez_as17}. Of particular interest are VCSELs with saturable absorption as external holding beam is not necessary  \cite{Fedorov_pre00,Bache_apb05,Genevet_prl08,Elsass_apb10,Kaur_jlt18}.

In this Letter, we introduce a spin-flip model for a broad-area VCSEL with a saturable absorber and  demonstrate a period doubling route to spatially localized chaos of elliptically polarized CSs. CS, which become chaotic by period doubling have been predicted for a driven damped nonlinear Shr\"{o}dinger equation \cite{Nozaki_pd86,Barashenkov_pre11}, for a Josephson junction ladder, for forced and damped van der Pol model \cite{Martinez_epl99} and for semiconductor laser with saturable absorber subject to optical feedback \cite{Panajotov_ol14,Schelte_pra17}. Oscillatory dynamics of localized structures has been experimentally observed in optically pumped VCSEL with saturable absorber \cite{Elsass_epjd10}. Experimental studies of devices based on liquid crystals have shown that nonvariational effects can give rise to chaotic behavior of multiple-peaks localized solutions \cite{Clerc-Residori_prl13}. In all these studies, polarization degree of freedom has not been considered. Recently, a coexistence of cavity solitons with different polarization states and different power peaks has been demonstrated for birefringent all-fiber resonators \cite{Averlant_ol17}.
To the best of our knowledge, the impact of light polarization on the CSs dynamics in a broad area semiconductor laser with saturable absorber has not been reported.

We consider the spin-flip VCSEL model in which we introduce saturable absorption leading to the following dimensionless partial differential equations
\begin{eqnarray}
\partial E_{x,y}/\partial t &=& [(1-i\alpha)N + (1-i\beta)N_a - 1\label{eq:dExydt} \\ &\mp& (\gamma_a+i\gamma_p) +  i\nabla^{2}_{\perp}]E_{x,y}\pm(i+\alpha)nE_{y,x},\nonumber \\
\partial N/\partial t &=& -b_1 [N(1 + |E_x|^2+|E_y|^2)\\ \nonumber &+& in(E_yE_x^*-E_xE_y^*) - \mu], \label{eq:dNxydt}\\
\partial n/\partial t &=& -b_sn-b_1[n(|E_x|^2+|E_y|^2)\\ \nonumber &+&iN(E_yE_x^*-E_xE_y^*)], \label{eq:dndt} \\
\partial N_a/\partial t &=& -b_2 [a_0 + N_a(1 + s|E_x|^2+|E_y|^2)]. \label{eq:dNadt_sfm}
\end{eqnarray}
Here $E_x$ ($E_y$) is the slowly varying mean electric field envelope for light polarized along the $x$ ($y$) direction, $N$ ($n$) is the sum  (difference) of the carrier densities of the two carrier sub-systems of the gain material with different projection of the spin and $\gamma_a$ and $\gamma_p$ represent the non-dimensional amplitude and phase anisotropies that couple the two field components (divided by the photon decay rate). $N_a$ is related to the carrier density in the absorber material with linear absorption $a_0$. $\alpha$ ($\beta$) is the linewidth enhancement factor and $b_1$  ($b_2$) is the ratio of photon lifetime to the carrier lifetime in the active layer (saturable absorber). $b_s$ is the nondimensional spin-flip rate, i.e. the ordinary SFM spin-flip rate as in \cite{SanMiguel_pra95} multiplied by the photon lifetime. $\mu$ is the normalized injection current in the active material, $a_0$ measures absorption in the passive material, and $s=a_2b_1/(a_1b_2)$ is the saturation parameter with $a_{1(2)}$ the differential gain of the active (absorptive) material. The diffraction of intracavity light $E$ is described by the Laplace operator $\nabla^{2}_{\perp}$ acting on the transverse plane $(x,y)$ and carrier diffusion and bimolecular recombination are neglected. Time and space are scaled to the photon lifetime $\tau_p$ and diffraction length.

Equations (\ref{eq:dExydt}) - (\ref{eq:dNadt_sfm}), admit a trivial solution: $E_{0x}=0$, $E_{0y}=0$, $N_0=\mu$,$n_0=0$, $N_{0a}=-a_0$, which is stable until laser threshold for linearly polarized (LP) emission, i.e. $\mu^{th}_{x/y}=1+a_0\pm \gamma_a$, and unstable afterwards. Nontrivial LP steady-state plane-wave solutions are obtained by substituting in eqns. (\ref{eq:dExydt})-(\ref{eq:dNadt_sfm}) the anzats $\{E_{x0},\, 0,\, N_{x0},\, 0,\, N_{ax0}\}$ for x-LP solution and $\{0,\, E_{y0},\, N_{y0},\, 0,\, N_{ay0}\}$ for y-LP solution. Here, $E_{x0,y0}=\sqrt{I_{x0,y0}}\,\mathrm{e}^{i\omega_{x,y}t}$, $N_{x0,y0}=\mu/(1+I_{x0,y0})$, $N_{ax0,ay0}=-a_0/(1+sI_{x0,y0})$, $n=0$, which gives quadratic equation for $I_{x0,y0}=E_{x0,y0}^2$ from
\begin{equation}
\frac{\mu}{1+I_{x0,y0}}-\frac{a_0}{1+sI_{x0,y0}}=1\pm \gamma_a.\label{eq:Istst}
\end{equation}
The frequencies $\omega_{x,y}$ of the steady state LP solutions depend on the intensity when the active and passive materials have different linewidth enhancement factors \cite{Bache_apb05} as $\omega_{x,y} = \mp\gamma_p-\alpha(1+\gamma_a)+ a_0\frac{\beta-\alpha}{1+sI_{x0,y0}}$. However, the stability of the lasing LP solutions does not depend on $\omega_{x,y}$. From eqn. (\ref{eq:Istst}) it follows that the LP light-vs-current (LI) characteristics have $C$ shape, i.e., turning points with $d\mu/dI{x,y}=0$ if
$s>1+(1\pm\gamma_a)/a_0$. The turning point coordinates in the LI plane are given by $I_{TPx,TPy}=[-1+\sqrt{a_0(s-1)/(1\pm\gamma_a)}]/s$ and  $\mu_{TPx,TPy} = [\sqrt{(s-1)(1\pm\gamma_a)}+\sqrt{a_0}]^2/s$.

\begin{figure}[t!]
\begin{picture}(60,240)(-25,0)
\put(-100,-10){\epsfxsize=240pt\epsfbox{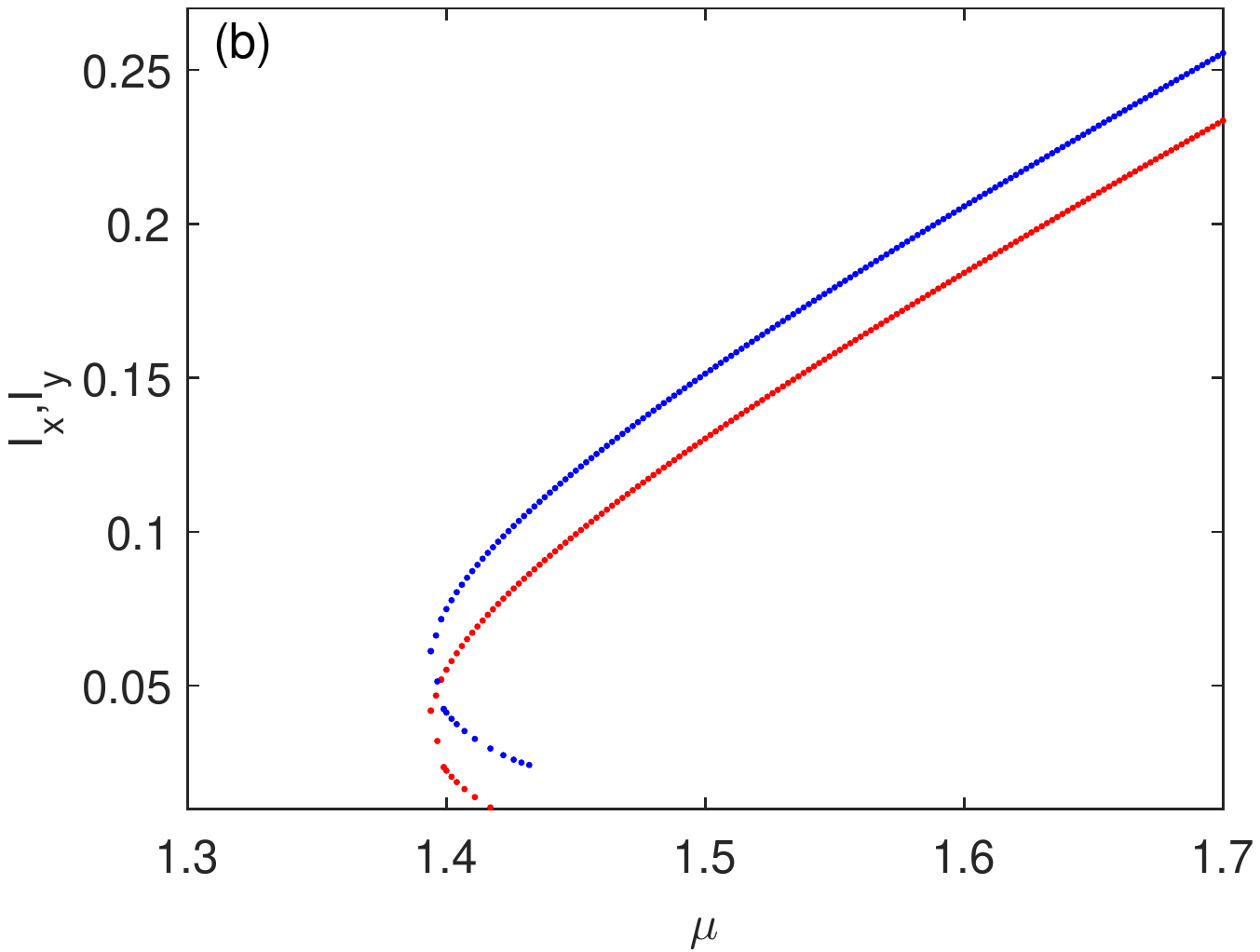}}
\put(-100,120){\epsfxsize=240pt\epsfbox{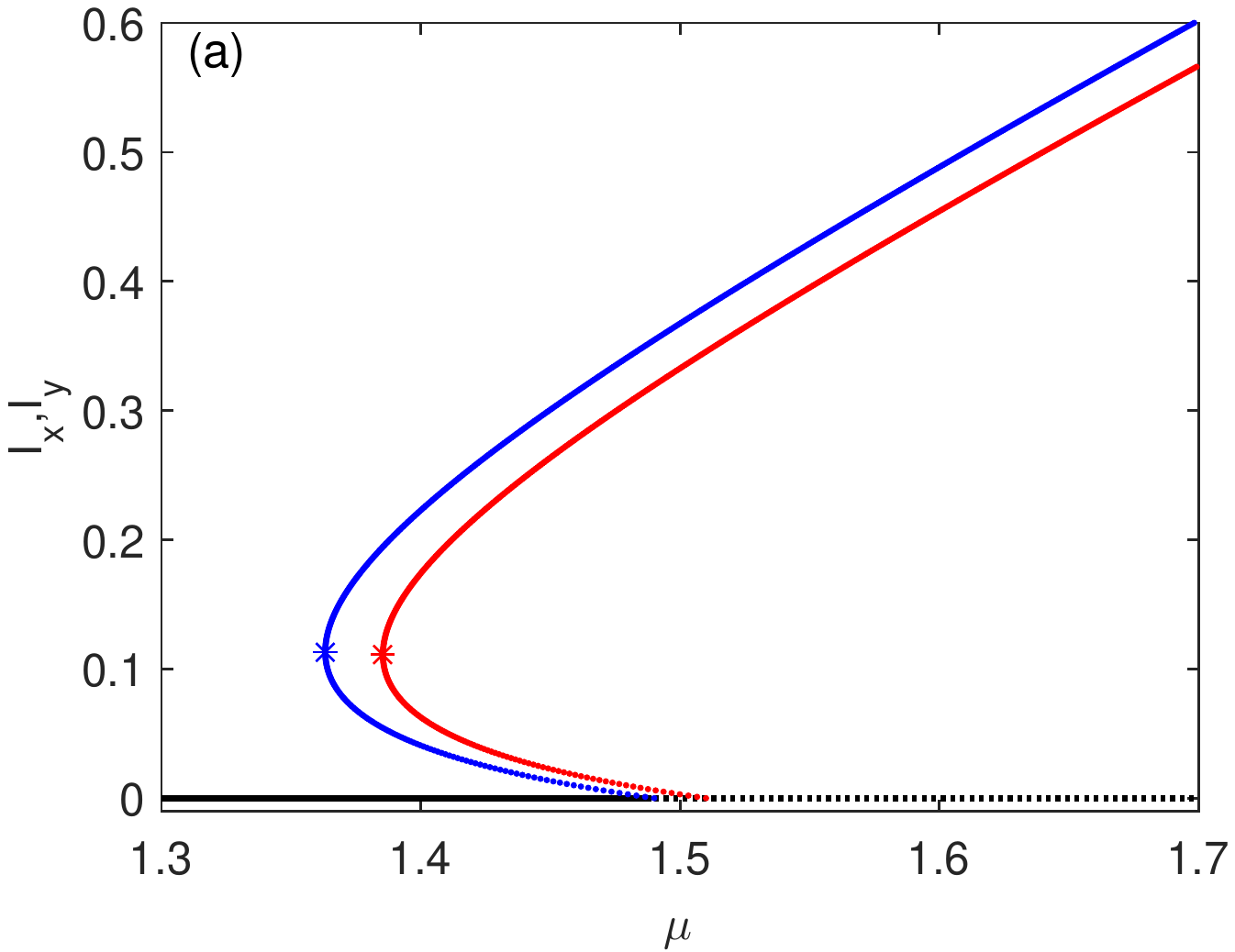}}
\end{picture}
\caption{(color on line) x (red) and y (blue) intensities of the steady state light-vs-current solutions of the SFM model with a saturable absorber: (a) LP solutions for $\gamma_a=0.01$ and $\gamma_p=0.01$. The turning points of the $C$ shaped LI curves are denoted by stars. The trivial (nonlasing solution is shown by solid (dotted) black line for stable (unstable) case. (b) Elliptically polarized solution for $\gamma_a=0.001$, $\gamma_p=0.01$.}
\label{fig:LI}
\end{figure}

The linear stability analysis of the LP stationary solutions is carried out by considering small fluctuations modulated with transverse wavenumber $K$ and leads to two independent systems, i.e. the x (y) LP solution can be destabilized by fluctuations of the orthogonal linear polarization $\delta E_y$ ($\delta E_x$) with simultaneous fluctuations in excess spin-polarized carries $n$ and/or by fluctuations in the same polarization $\delta E_x$ ($\delta E_y$) accompanied by fluctuations in the overall carrier density $\delta N$ and saturable absorber carrier density $\delta N_a$. An example of LP light vs current (LI) characteristics is presented in Fig. \ref{fig:LI} (a). {Here, and in the following we consider the laser parameters:} $\mu=1.45$, $\alpha=2$, $\beta=0$, $b_1=0.04$, $b_2=0.02$, $b_s=0.25$, $a_0=0.5$, $s=10$. The negative branch of the LI curve is plane wave (i.e. $K=0$) unstable with one positive real eigenvalue and two complex conjugated eigenvalues with negative real part. The positive branch of the $C$ shape LI curve is plane wave stable for the parameter choice. However, a Hopf bifurcation ($\lambda =i\omega$) can take place on this branch for an increased ratio $b_2/b_1$ \cite{Bache_apb05}.

The nontrivial elliptically polarized (EP) steady-state plane-wave solutions are obtained by substituting in eqns. (\ref{eq:dExydt})-(\ref{eq:dNadt_sfm}) the anzats $\{E_{ex}e^{i\omega_e t},\, E_{ey}e^{i\omega_e t},\, N_e,\, n_e,\, N_{ae}\}$, which leads to a system of five nonlinear algebraic equations. This system can only be solved numerically, the obtained EP steady-state solutions for $\gamma_a=0.001$ and $\gamma_p=0.01$ are shown in Fig. \ref{fig:LI} (b).

\begin{figure}[h!]
\begin{picture}(60,220)(-25,0)
\put(-130,-5){\epsfxsize=220pt\epsfbox{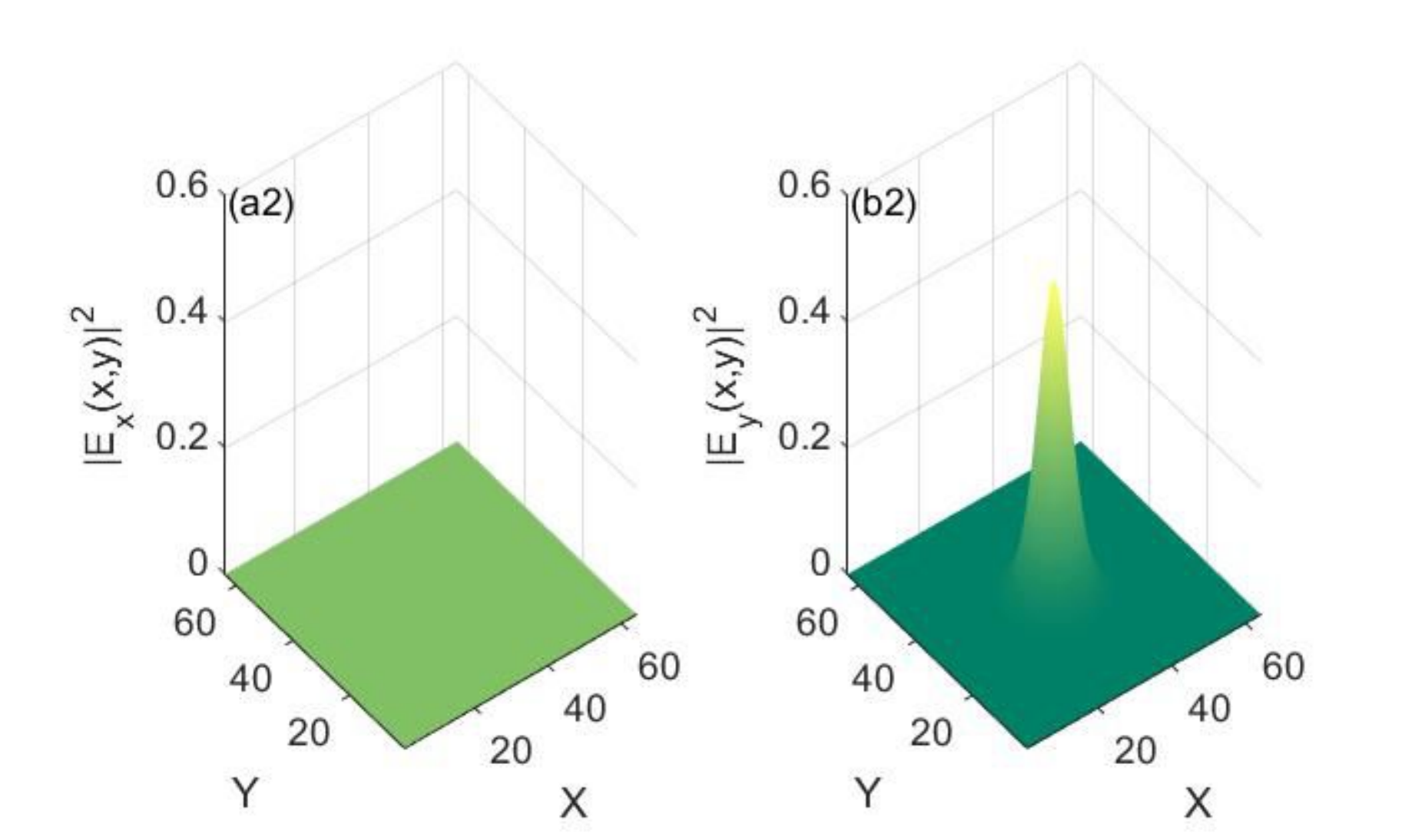}}
\put(-130,110){\epsfxsize=220pt\epsfbox{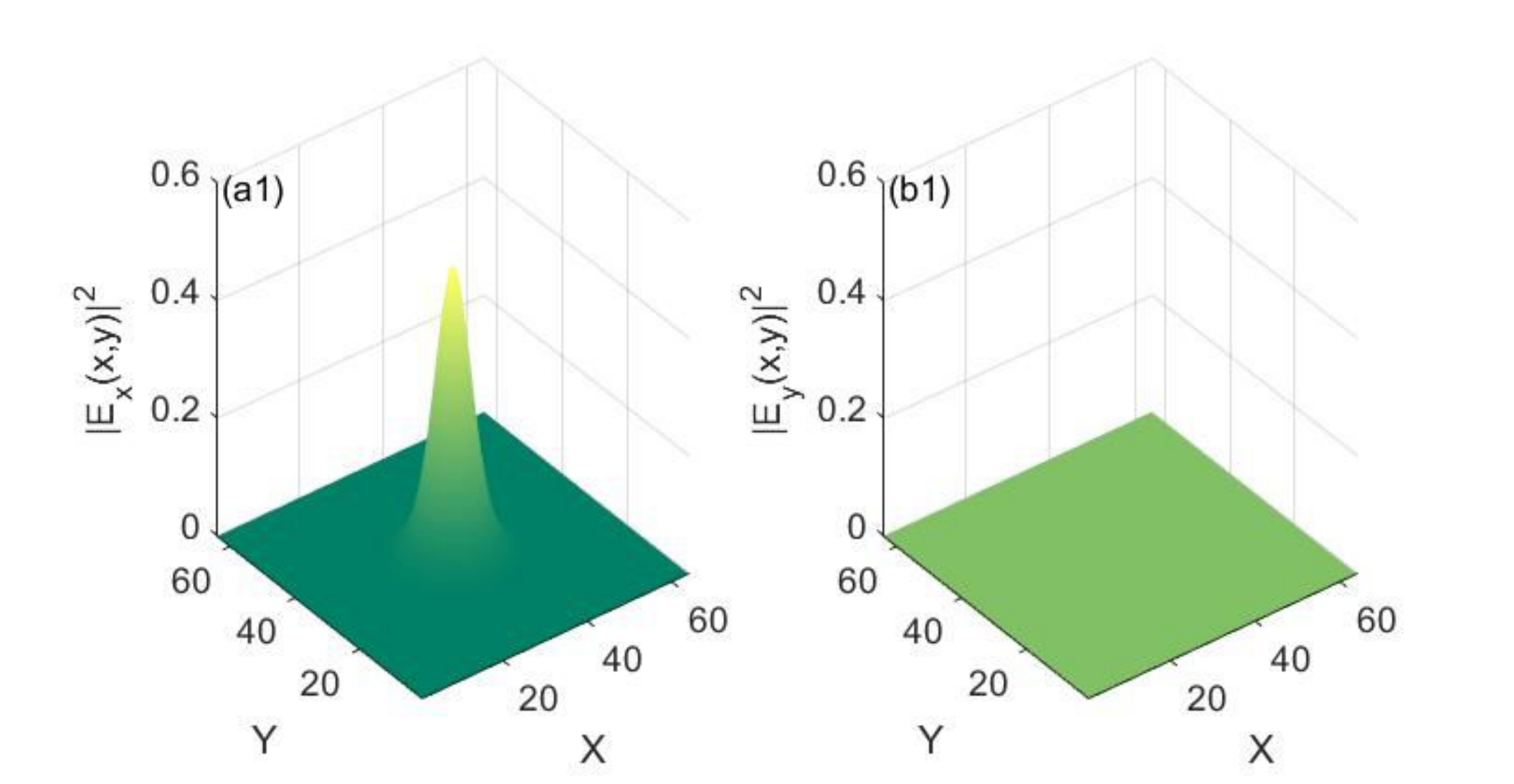}}
\end{picture}
\caption{Linearly polarized cavity solitons for $\gamma_p=0.05$ and $\gamma_a=0.001$: (a1),(b1) x LP CS and (a2),(b2) y LP CS.}
\label{fig:CS}
\end{figure}

\begin{figure}[t!]
\begin{picture}(60,210)(-25,0)
\put(-120,-10){\epsfxsize=220pt\epsfbox{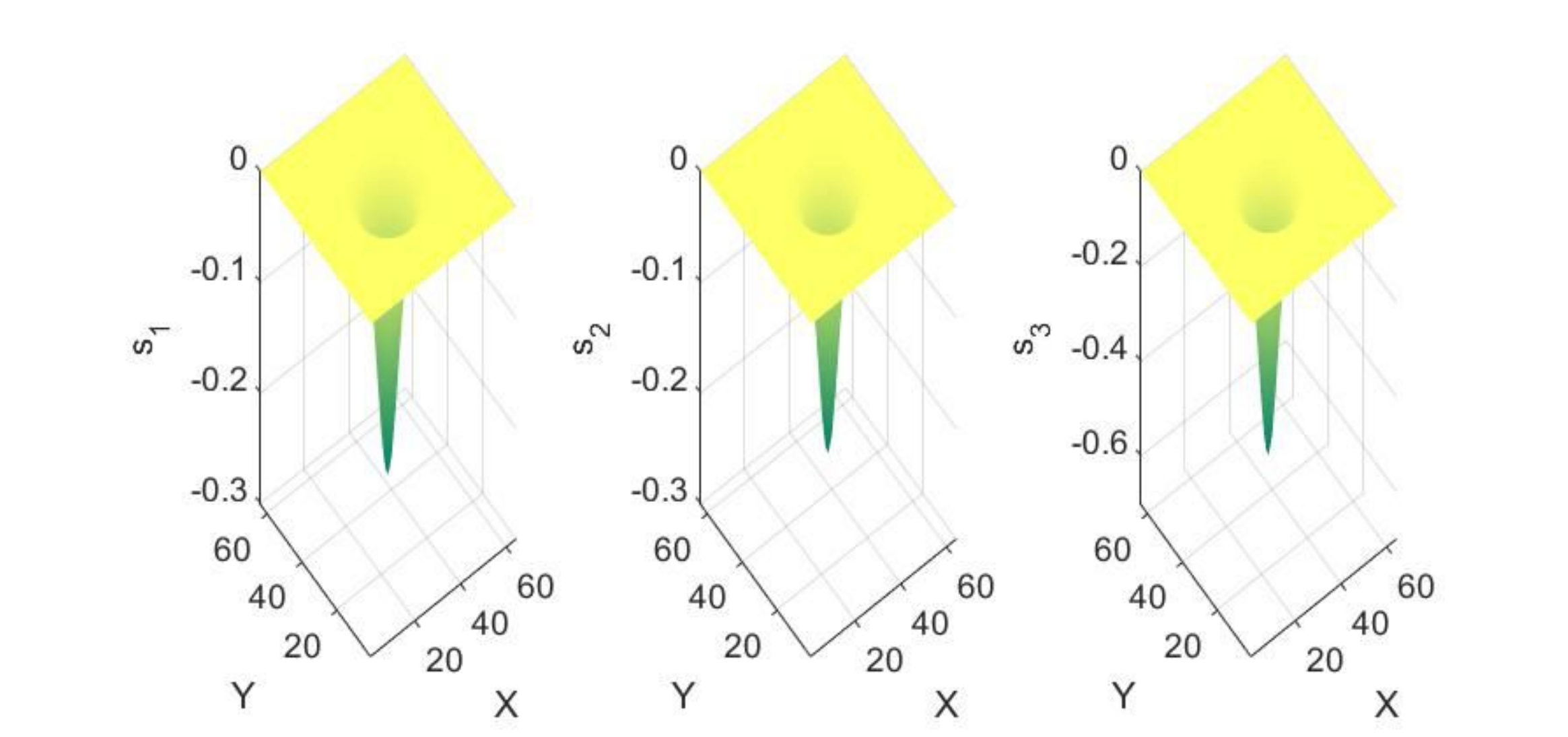}}
\put(-120,80){$(b)$}
\put(-120,100){\epsfxsize=220pt\epsfbox{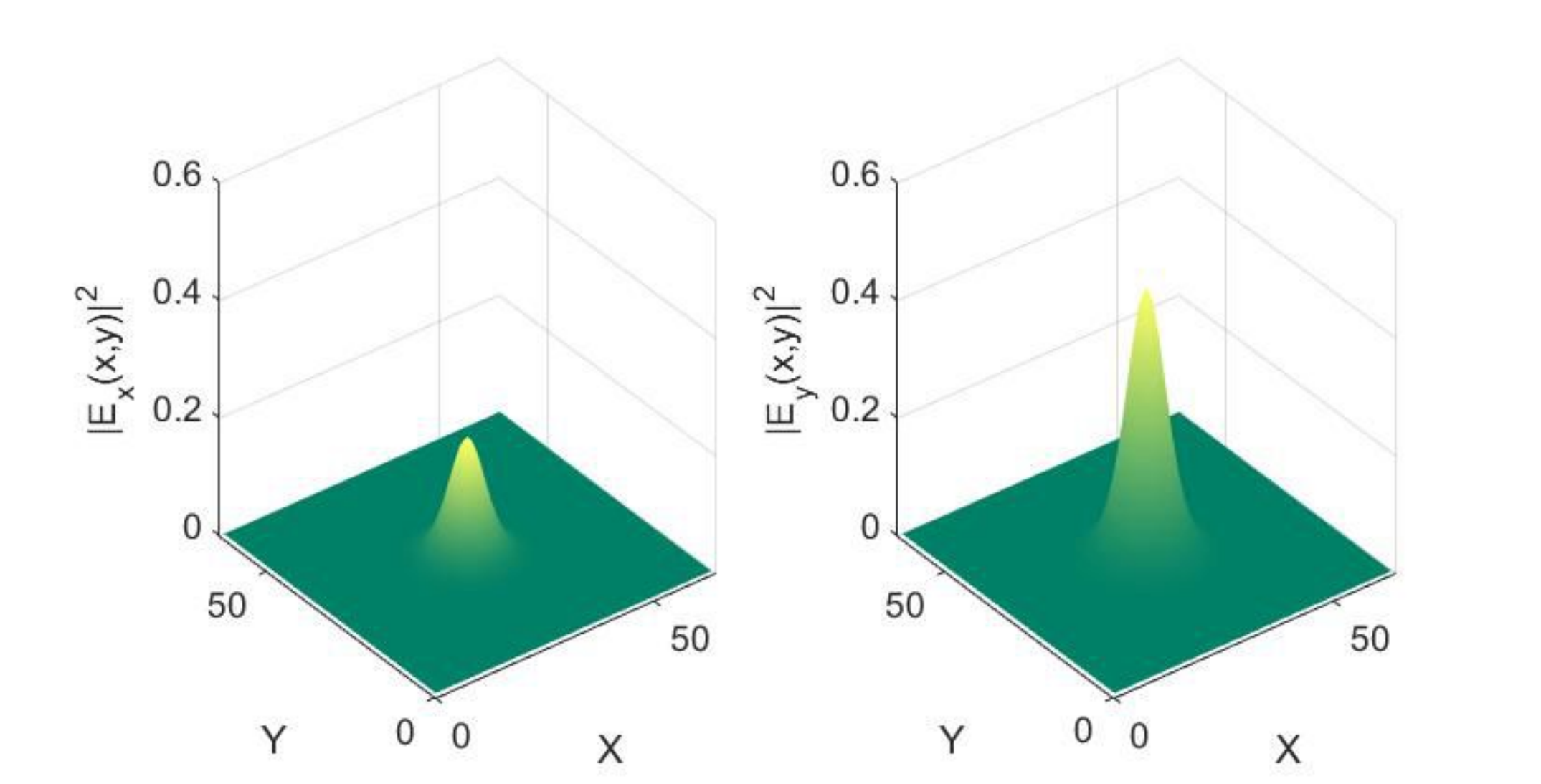}}
\put(-120,180){$(a)$}
\end{picture}
\caption{Elliptically polarized cavity soliton for the same parameters as in Fig. \ref{fig:CS} and $t=730$. (a) x and y intensity distribution; (b) spatial distribution of Stokes parameters.}
\label{fig:ellipCS}
\end{figure}

For our choice of parameters the upper branches of the LI curves exhibit Turing bifurcation allowing for the formation of CSs. Due to the steady state multistability of the system, we expect it to also exhibit CS polarization multistability. This is indeed confirmed in Figs. \ref{fig:CS} and \ref{fig:ellipCS}, which are obtained by integrating numerically Eqs. (\ref{eq:dExydt}) - (\ref{eq:dNadt_sfm}) by the split-step method with periodic boundary conditions for $\gamma_p=0.05$ and $\gamma_a=0.001$. Fig. \ref{fig:CS} presents the case of linearly polarized CS: x-LP (Fig. \ref{fig:CS} a1,b1) and y-LP (Fig. \ref{fig:CS} a2,b2), which exist in the same region of injection currents as for the scalar case \cite{Bache_apb05}: $\mu \in [1.42, 1.49]$. Fig. \ref{fig:ellipCS} presents the case of an elliptically polarized CS: in (a) the intensities of x and y polarization and in (b) the Stokes parameters for at time $t = 730$. This figure reveals the profound difference to the LP case and establishes that it is indeed a CS with elliptical polarization. Another striking difference to the LP CSs is that the EP CS does not reach a stable state but persists in a complex temporal dynamics state. The injection current region of existence of EP CS is smaller than the one for LP CS: $\mu \in [1.43, 1.48]$.

\begin{figure}[h!]
\begin{picture}(60,150)(-25,0)
\put(-120,-10){\epsfxsize=320pt\epsfbox{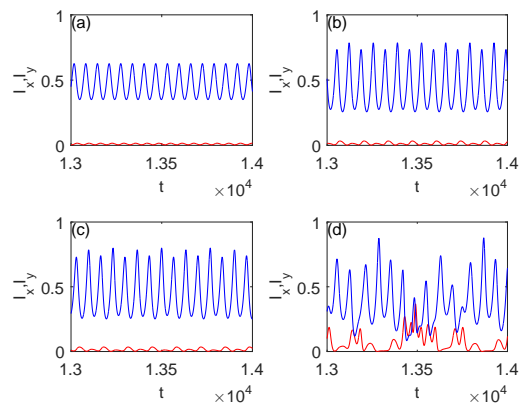}}
\end{picture}
\caption{(color online) Time traces of the CS peak intensity (x-LP with red color and y-LP with blue color) for different values of the phase anisotropy revealing different dynamics: (a) period one for $\gamma_p = 0.07$, (b)  period two for $\gamma_p = 0.06$, (c) period four for $\gamma_p = 0.0596$ and (d) chaos for $\gamma_p = 0.05$.}
\label{fig:CS_t}
\end{figure}

\begin{figure}[h!]
\begin{picture}(60,150)(-25,0)
\put(-120,-10){\epsfxsize=320pt\epsfbox{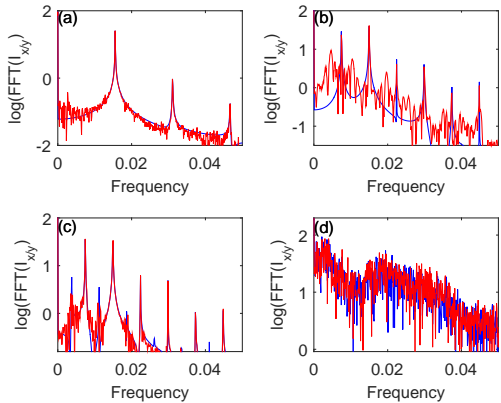}}
\end{picture}
\caption{(color online) Power spectra corresponding to the CS peak time traces of Fig. \ref{fig:CS_t} are shown with blue color. The spectra with the red color correspond to the case when spontaneous emission is accounted as a white noise term with a strength of $1\times10^{-4}$.}
\label{fig:PowerSpectra}
\end{figure}

In order to characterize the temporal dynamics of the EP CS, we fix all parameters mentioned above, and vary the phase anisotropy $\gamma_p$. For $\gamma_p = 0.07$, the CS exhibits regular time oscillations and displays period one dynamics (P1) as shown in  Fig. \ref{fig:CS_t}a. The frequency of the oscillations is measured to be $\nu\approx0.014$, which is close to the phase anisotropy splitting of $\gamma_p/2\pi=0.011$. Decreasing $\gamma_p$, brings the CS to period two dynamics (P2) (Fig. \ref{fig:CS_t}b) and then to period four dynamics (Fig. \ref{fig:CS_t}c). Finally, Fig. \ref{fig:CS_t}d presents an example of chaotic localized dynamics, i.e. the CS polarization dynamics is a period-doubling route chaos. The transition between CS of different height is continuous and smooth and the CS does not spread considerably as its peak amplitude changes in time. Period doubling route to polarization chaos has been observed in small-area single-transverse mode solitary VCSELs in only two circumstances: with quantum dots as an active material \cite{Virte_np13} and with quantum wells as an active material but subjected to external anisotropic strain \cite{Raddo_sr17}. The system considered here differs in two ways, first, the VCSEL is a broad-area, highly multi-transverse-mode device and second, there is a saturable absorber integrated within the laser. Indeed, such a laser allows for the existence of stable cavity solitons, which undergo the period doubling cascade to chaos reported here.
Fig. \ref{fig:PowerSpectra} shows the corresponding power spectra to Fig. \ref{fig:CS_t}: the blue (red) color is without (with) spontaneous emission noise. In the case of P1 dynamics, the power spectrum reveals strong peak at $f_1 = 0.014$ and multiples of it (Fig. \ref{fig:PowerSpectra}a). For the case of P2 dynamics, additional peaks appear at frequencies of $f_1/2$ and multiples (Fig. \ref{fig:PowerSpectra}b). RF spectrum in a regime of period quadrupling is shown in Fig. \ref{fig:PowerSpectra}c for $\gamma_p=0.0596$.  Finally, for phase anisotropy of $\gamma_p=0.05$ when the CS peak amplitude changes chaotically the powers spectrum looses the sharp individual peaks features and becomes quite broad (Fig. \ref{fig:PowerSpectra}d).
We have checked that the main features of the reported dynamics are robust to noise. For example, in the case of Fig. \ref{fig:CS_t} (a)-(d), adding to Eq.~(1) a noise source with a mean value of zero and a strength of $1\times10^{-4}$ preserves the main features of the dynamics despite that the peaks heights in the time traces become with somewhat varying amplitude. Thus, the spectra reported in Fig. \ref{fig:PowerSpectra} (red color) show quite similar peaks for the P1 dynamics and somewhat blurred P2 and P4 dynamics features with increased background with respect to the noiseless case (blue color).
\begin{figure}[h!]
\begin{picture}(60,150)(-25,0)
\put(-120,-10){\epsfxsize=320pt\epsfbox{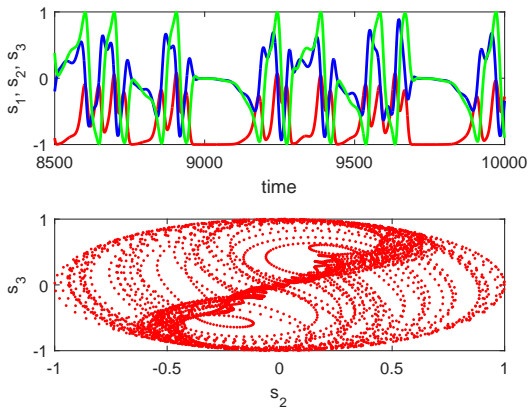}}
\end{picture}
\caption{(color online) (a) Time traces of the Stokes parameters for the CS peak $s_1$ - red, $s_2$ - blue and $s_3$ - green. (b) the corresponding trajectory in a cross section of the phase space spanned by the Stokes parameter $s_2$ and $s_3$. The CS is in the chaotic temporal dynamics region: $\gamma_p = 0.05$.}
\label{fig:Chaos_time}
\end{figure}
Time traces of the Stokes parameters for the CS peak are presented in Fig. \ref{fig:Chaos_time}a and the corresponding trajectory in a cross section of the phase space spanned by the Stokes parameter $s_2$ and $s_3$ in Fig. \ref{fig:Chaos_time}b. Although the laser dwells most of the time around two well separated states of $s2$ and $s3$ of opposite sign,the trajectory goes well beyond these state filling in the entire plane.

In conclusion, we have introduced a spin-flip model for a broad-area VCSEL with a saturable absorber and demonstrated coexistence of linearly polarized (either along the x or the y axis) and elliptically polarized (EP) cavity solitons. Depending on the phase anisotropy of the VCSEL, which accounts for the birefringence splitting in the laser, a period doubling route to chaos of a single spatially localized light structure is observed. The phase trajectory makes a wide excursion in the Stokes parameters space when the localized structure is in a state of chaotic temporal dynamics. These findings expand the numerous work on CSs in VCSELs, which so far consider mostly the scalar rather than vectorial case (light polarization is neglected).  Our work not only provides a way to consider vector CS in a VCSEL with saturable absorber, but broadens their possible applications as bits of information (multilevel logic due to the CS polarization multistability) or as two-dimensional arrays of re-writable laser-source with non-trivial temporal dynamics.

K.P. acknowledges the Methusalem foundation for financial support. M.T. is a Research Director with the Fonds de la Recherche Scientifique F.R.S.-FNRS, Belgium.

\end{document}